\documentclass[referee]{raa}
\usepackage{graphicx,times}
\usepackage{natbib}
\usepackage{amssymb,amsmath}
\bibpunct{(}{)}{;}{a}{}{,}

\usepackage[a4paper=true,dvipdfm=true,pagebackref=true]{hyperref}
\hypersetup{pdftitle = The title of my PDF, pdfauthor = My name, pdfsubject= The subject, pdfkeywords = keyword1 keyword2 keyword3}
\hypersetup{colorlinks = true, linkcolor = green, anchorcolor = red, citecolor = blue, filecolor = red, pagecolor = red, urlcolor = red}
\begin{document}

\title{Correcting for the solar wind in pulsar timing observations: the role of simultaneous and low-frequency observations
}

   \volnopage{Vol.0 (200x) No.0, 000--000}      
   \setcounter{page}{1}          

   \author{Ze-xi Niu
      \inst{1}
   \and George Hobbs
      \inst{2}
   \and Jing-bo Wang
      \inst{3,4}
   \and Shi Dai
      \inst{2}}

   \institute{Department Of Astronomy, Beijing Normal University,
             Beijing 100875, China; \\
        \and
             Australia Telescope National Facility, CSIRO, PO Box 76 Epping, NSW 1710, Australia\\
        \and
             Xinjiang Astronomical Observatory, Chinese Academy of Sciences, 150 Science 1-Street, Urumqi, Xinjiang 830011, China; {\it wangjingbo@xao.ac.cn}
        \and
Key Laboratory of Radio Astronomy, Chinese Academy of Sciences, 150 Science 1-Street, Urumqi, Xinjiang, China, 830011 \\}
   \date{Received~~2009 month day; accepted~~2009~~month day}

\abstract{The primary goal of the pulsar timing array projects is to detect ultra-low-frequency gravitational waves.  The pulsar data sets are affected by numerous noise processes including varying dispersive delays in the interstellar medium and from the solar wind.  The solar wind can lead to rapidly changing variations that, with existing telescopes, can be hard to measure and then remove.  In this paper we study the possibility of using a low frequency telescope to aid in such correction for the Parkes Pulsar Timing Array (PPTA) and also discuss whether the ultra-wide-bandwidth receiver for the FAST telescope is sufficient to model the solar wind variations.  Our key result is that a single wide-bandwidth receiver can be used to model and remove the effect of the solar wind.  However, for pulsars that pass close to the Sun such as PSR~J1022$+$1022, the solar wind is so variable that observations at two telescopes separated by a day are insufficient to correct the solar wind effect.
}

   \authorrunning{Ze-xi Niu, George Hobbs, Jing-bo Wang \& Shi Dai}            
   \titlerunning{Correcting for the solar wind in pulsar timing observations}  

   \maketitle

\section{Introduction}           


Pulsar timing arrays (PTAs) projects, such as the Parkes Pulsar Timing Array\footnote{The Parkes Pulsar Timing Array (PPTA) in Australia is an implementation of the PTA concept based on observations with the Parkes 64-m radio telescope. 24 millisecond pulsars is being observed at three radio-frequency bands, 50 \,cm ($\sim$ 700 \,MHz), 20 \,cm ($\sim$ 1400 \,MHz) and 10 \,cm ($\sim$ 3100 \,MHz).}\citep{2013PASA...30...17M}, aim to detect ultra-low-frequency gravitational waves by observing a large number of millisecond pulsars (MSPs) \citep{2015Sci...349.1522S}. To do this, these pulsars are observed every few weeks over decades. Differences between the measured and predicted pulse times of arrival (ToAs) are referred to as ``timing residuals" \citep{2006MNRAS.369..655H}. Gravitational waves are usually not included in the timing model and therefore the existence of such waves will mean that the ToAs cannot be perfectly predicted with the model and hence gravitational waves will induce timing residuals.   Other phenomena can also lead to discrepancies between the measurements and the predictions. For instance, the timing model may be incorrect, the clock used for determining the ToAs may have an error and the pulsars themselves may not be perfectly stable \citep{2004MNRAS.353.1311H}.  In many cases the dominant noise process are variations in a pulsar's dispersion measure (DM) caused by the fluctuations in the interstellar medium or in the solar wind \citep{2007ApJ...671..907Y}. The challenge for the PTA experiments is therefore to identify the signature caused by the gravitational wave signal in the presence of these other noise processes. This can be done by looking for correlations between the timing residuals for different pulsars \citep{2015Sci...349.1522S}. However, \cite{2016MNRAS.455.4339T} showed that, for a realistic array of pulsars, phenomena such as errors in the solar system ephemeris, instrumental errors or uncorrected solar wind effects potentially can lead to false detections of the gravitational wave signal.

The dispersive group delay for a pulsar signal, $t_{\rm DM}$, is related to the integral of the electron density, $n_e$, from the pulsar to the Earth:
\begin{equation}
t_{\rm DM} = \lambda^2 \left(\frac{e^2}{2 m_e \pi c^3}\int_{0}^{d} n_e \,dl\right)
\end{equation}
where ${\rm DM} = \int_{0}^{d} n_e \,dl$ is the `dispersion measure (DM)', $\lambda$ is the observing wavelength, $c$ is the speed of light, $e$ is the electron charge, $m_e$ is the mass of an electron and $d$ is the distance between the Earth and the pulsar. The DM at a given time can be determined if pulse arrival times have been measured at multiple observing frequencies.   The precision with which the DM can be measured is determined by the arrival time uncertainties and the frequency coverage of the observations. If the DM variations can be measured with sufficient precision then such variations can be accounted for in the pulsar timing model.  Methods to do this for the PPTA sample of pulsars were described by \cite{2007MNRAS.378..493Y}, \cite{2013MNRAS.429.2161K} and \cite{2016MNRAS.455.1751R}\footnote{PTA experiments in Europe and in North America have also studied DM variations; see \cite{2016MNRAS.458.3341D} and \cite{2016ApJ...821...66L}.}.  These papers showed that the DM variations can often, but not always, be modelled by assuming Kolmogorov turbulence in the interstellar medium.  However, \cite{2013MNRAS.429.2161K} showed that, without care, the method used to account for DM variations can remove an underlying gravitational wave signal.  By simultaneously modelling the frequency-dependent signal (the DM variations) along with a frequency-independent signal (known as the ``common mode" and would include the gravitational wave signal), it was shown that the gravitational wave signal would not be affected.  This method works well when the DM variations are relatively smooth and the changes occur over the timescale of many observations (i.e., weeks to months).

\cite{2007ApJ...671..907Y} and \cite{2012MNRAS.422.1160Y} studied observations of PPTA pulsars whose line-of-sight pass close to the Sun. They demonstrated that, at an observing frequency of 1400\,MHz, the solar wind can contribute ToA errors of 100 ns for sources 60$^\circ$ from the Sun and more than 1$\mu s$ within 7$^\circ$. In contrast to the variations caused by the turbulence in the interstellar medium, the DM variations caused by the solar wind depend strongly on the line-of-sight angle to the Sun and change on various time-scales including day-to-day variability and multi-year variability because of the solar cycle.

The standard pulsar timing software packages such as \textsc{tempo} and \textsc{tempo2} \citep{2006MNRAS.369..655H} include simple models to account for the solar wind. The default model is a time-independent, spherically symmetric model completely defined by the angle of the line-of-sight to the Sun and a single electron density parameter (NE1AU: the electron density at 1\,A.U.). \cite{2007MNRAS.378..493Y} demonstrated that this parameterisation was not suitable for modelling the actual observations\footnote{As part of our work, we updated this model to allow changes in the electron density parameter as a function of time, but we determined that even this model was too simplistic to model real changes in the solar wind.}. \cite{2007MNRAS.378..493Y} attempted to model the measured DM changes using observations from the Wilcox Solar Observatory\footnote{The Wilcox Solar Observatory (WSO), located in California in US has been collecting solar polar field data since 1975. The solar field data can be used to predict the magnitude of the next cycle and the peak of the current cycle.}. The WSO models do predict variations in the DM changes caused by the solar wind that are similar to what we observe, but as shown by \cite{2007MNRAS.378..493Y}, they are not sufficient for correcting a given pulsar data set.

To date, the PPTA observations have been carried out either with a relatively small bandwidth centred in the 20\,cm observing band, or using a dual-band (10/50 \,cm) receiver. In a typical observing session, a given pulsar is first observed using one of these receivers and then (perhaps a few days later) observed using the other receiver. Although this is not regularly done, we can also obtain estimates of a pulsar's DM at a given time using low frequency telescopes, such as the Murchison Widefield Array (MWA) \citep{2014ApJ...791L..32B}\footnote{The Murchison Widefield Array (MWA) is a Square Kilometre Array (SKA; \citealt{2009IEEEP..97.1482D}) precursor telescope at low radio frequencies in western Australia.}. This provides very low frequency (80-300\,MHz) observations, but not necessarily at the same time as the Parkes observations. We are also currently commissioning an ultra-wide-bandwidth receiver for Parkes that should enable us to simultaneously observe from 700\,MHz to 3GHz.  Many other observatories are also building ultra-wide-bandwidth receivers. For instance, the Five-hundred-meter Aperture Spherical Telescope\footnote{The Five-hundred-meter Aperture Spherical Telescope (FAST), situated is GuiZhou province in China, is the world's largest single dish radio telescope and opened on 25th September 2016.} will have a receiver providing a band from 270\,MHz to 1620\,MHz \citep{2016RAA....16..151Z}.

In this paper, we:
\begin{itemize}
\item demonstrate how realistic solar wind effects can be simulated within the \textsc{tempo2} software package
\item simulate (1) realistic observations for the ultra-wide-band receiver with the FAST telescope and (2) observations in the low-frequency band with MWA and in the 20\,cm band with Parkes telescope.
\item trial different methods for measuring and removing the DM variations and test whether any underlying gravitational wave signal would be affected.
\item discuss the implications of these results for FAST and the PPTA.
\end{itemize}

\section{Simulation}

We used the \textsc{ptaSimulate} software package to simulate pulsar parameters and ToAs. This software package was developed in order to make realistic simulations of pulsar timing array data sets. For our work, we choose only to make a simple simulation of the radiometer noise (leading to the errors on the ToAs), but realistic simulations of an isotropic, stochastic gravitational waves and of the solar wind. For the radiometer noise we simply assume that the ToA uncertainties are 100 ns which is typical of a ``good" pulsar in the PPTA. For the gravitational waves we use the routines described by \cite{2009MNRAS.394.1945H} to simulate a background defined by:
\begin{equation}
h = \left(\frac{A^2}{12{\pi}^2}\right){\left(\frac{f}{f_y}\right)}^{{2\alpha}-3}
\end{equation}
where $\alpha = -2/3$, $f_y = 1/{\rm 1 yr}$ and we choose $A = 10^{-14}$. This amplitude is much larger than our current upper bounds \citep{2015Sci...349.1522S}, but is used to demonstrate the possibility that the gravitational wave signal can be affected by the correction method used to model the solar wind. The solar wind is modelled based on the \textsc{tempo2} implementation described in \cite{2007ApJ...671..907Y} which uses WSO data as input. We have simulated various pulsars in the PPTA sample, but for this work, we concentrate on PSR~J1022$+$1001 which has the lowest ecliptic latitude ($-0.06^\circ$) of the pulsars in the sample. We use \textsc{ptaSimulate} to provide three noise models: (N1) includes the radiometer noise and gravitational waves, (N2) the radiometer noise and the solar wind and (N3) all three noise processes.

\begin{figure}[htb]
\centering
\includegraphics[angle=270,width=11cm]{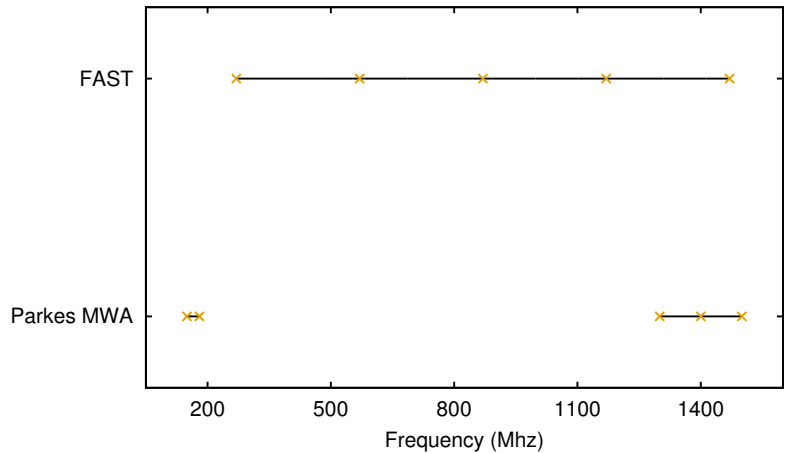}
\caption{The simulated frequency range for each telescope. Black bars indicate the simulated frequency range and the orange crosses indicate the exact frequencies used in the simulation.}
\label{Fig1}
\end{figure}

For each of the noise processes we make two data sets:
\begin{itemize}
\item (DS1): we simulate the FAST telescope and the ultra-wide-band receiver covering a range from 270 to 1620 \,MHz and we assume that we obtain a single ToA for every 300\,MHz interval.
\item (DS2): we simulate the Parkes telescope observing in the 20 \,cm and the MWA telescope observing in lower band.
\end{itemize}

As shown in Figure 1, for DS1, we simulate 270, 570, 870, 1170 and 1470 \,MHz as suggested by Table 2 in \cite{2016RAA....16..151Z}. We assume that we obtain a single ToA for every 300 MHz interval\footnote{Pulsar timing has not yet been carried out with such a wide instantaneous bandwidth and so it is not clear exactly how the band will be divided.  We note that the North American Nanohertz Observatory for Gravitational Waves (NANOGrav) project \citep{2016ApJ...821...13A} obtain a ToA for each 4MHz subband. However, processing such a large number of ToAs is impractical with the computing resources that we have available and therefore we simulated ToAs every 300\,MHz.}. For DS2, we simulate 150 and 180\,MHz for MWA according to the Table 1 in \cite{2013PASA...30....7T} and Table1 in \cite{2015RaSc...50..574L} and 1300,1400 and 1500\,MHz for Parkes according to the Table 1 in \cite{2013PASA...30...17M}.

Even though the Parkes and MWA observations provide a frequency span from 80 to 3100 \,MHz, it is not common for both telescopes to observe simultaneously. Therefore, for DS2 we have also tried different time offsets between the MWA and the Parkes observations: (TO1) the observations were simultaneous; (TO2) the MWA observation occurred one day after the Parkes observations.  All of our simulated data sets have a data span of ten years and an observing cadence of 14\,d. We simulated from 2005 to 2015 as we use real WSO data sets for the Solar wind modelling - clearly the actual FAST and Parkes/MWA data do not cover this time range, but the results will not depend upon the exact dates chosen.

\begin{figure}[h]
\setlength{\belowcaptionskip}{-4.cm}
\centering
\includegraphics[angle=270,width=11cm]{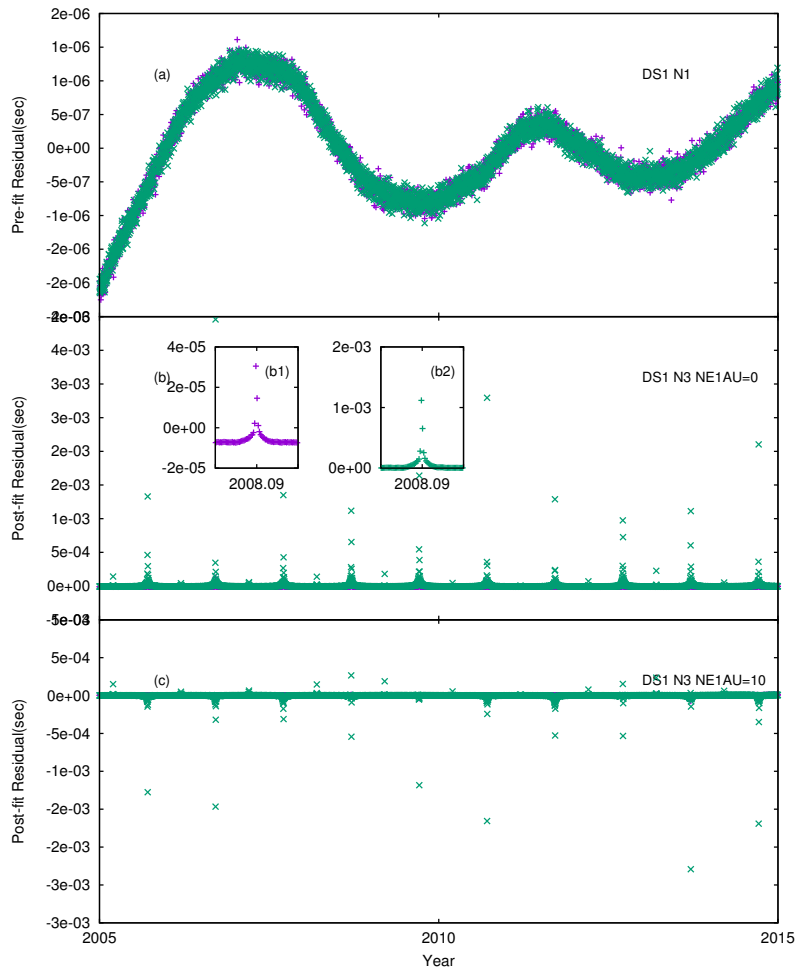}
\caption{Simulated residuals for FAST. Purple points mark simulated frequency 1470 \,MHz, green points mark simulated frequency 270 \,MHz. The top panel (a) shows expected white, radiometer noise and the low frequency gravitational wave signal.   The central panel (b) shows the radiometer noise and the effect of the solar wind. Panel (c) shows the same, but after attempting to model the solar wind using a simple, spherically-symmetric model.}
\label{Fig2}
\end{figure}

In order to illustrate the effect of the solar wind we have also simulated the highest and lowest (270\,MHz 1470\,MHz) observing bands for FAST, but with a 1 day sampling.  In Figure 2a we show a typical realisation of the gravitational wave background signal over the 10 year data span. In Figure 2b we show the simulated signal with the gravitational wave background and the solar wind. The solar wind clearly produces timing residuals many orders of magnitude larger than the gravitational wave signal.  As expected the signal has an annual periodicity representing the times when the line-of-sight to the pulsar goes close to the Sun.  The exact signal depends upon the solar cycle and the exact data sampling.  In the inset (b1) and (b2) to Figure~2b we show a zoom-in around year 2008 for 1470 \,MHz and 270 \,MHz.  In this panel we have not attempted to remove the effect of the solar wind using any method.  In Figure 2c we show the same data, but, this time, use the standard \textsc{tempo/tempo2} spherically symmetric solar wind model in which we assume that the electron density at 1AU is $10$\,e\,cm$^{-3}$. Clearly, significant (both positive and negative) residuals are still seen and so this simple model both over-corrects and under-corrects the data at different epochs and does not provide a sufficient means to remove the solar wind variations.

\section{Method and Results}

\begin{figure}[h]
\centering
\includegraphics[angle=270,width=9cm]{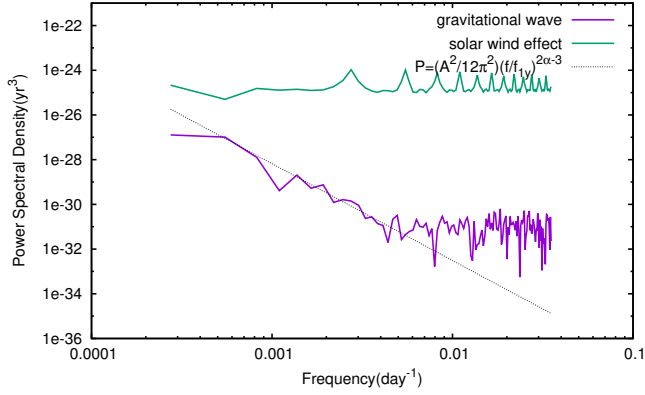}
\caption{Power spectral density of pre-fit residuals. The green line indicates the simulated solar wind effect and radiometer noise. The purple line marks the simulated gravitational wave and radiometer noise. The dashed line is the simulated power low model of the gravitational wave background signal.}
\label{Fig3}
\end{figure}

In Figure 3, we show the power spectral density of the pre-fit residuals for DS2. The green line shows the spectral density for the timing residuals affected by the solar wind and radiometer noise.  The purple line is the same, but for the simulation containing only the gravitational wave and radiometer noise.  The dashed line is the analytic representation of the power spectrum of the simulated gravitational waves. Producing a single spectrum using an identical method is challenging in this case as the gravitational wave signal has a steep power-law signal which requires that a pre-whitening method to mitigate the effects of spectral leakage. However, the solar wind signal does not follow a steep power-law spectrum and therefore we should not use pre-whitening methods.  Our analysis here is to remove the effect of the solar wind without removing the gravitational wave signal. We therefore present these power spectra in a way that is optimal for the steep gravitational wave power law spectrum and note that the spectra of the solar wind will not be perfect (particularly at low frequencies). Throughout this paper we therefore input a set of pulsar timing residuals and form a power spectrum assuming a gravitational wave background power law spectrum using the ``analyticChol" and ``cholSpectra" \textsc{tempo2} plugins\footnote{These two plugins use Cholesky fitting routines to obtain a power spectral of the timing residuals}. Note that this same procedure was followed by \cite{2016MNRAS.455.4339T}. Clearly, without correction the solar wind effect dominates the entire spectrum and makes the gravitational waves undetectable. Our aim in this section is to determine whether existing methods can be used to remove the effect of the solar wind, without also removing the gravitational wave signal.

\begin{figure}[h]
\centering
\includegraphics[angle=270,width=11cm]{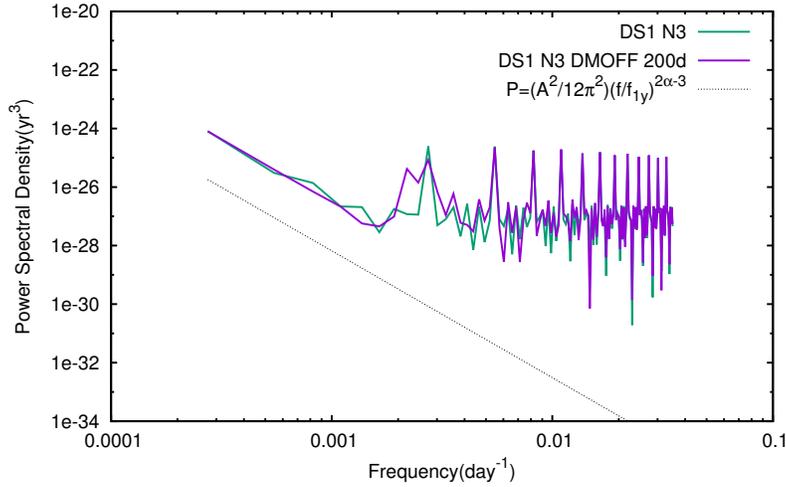}
\caption{Power spectral density for the simulated N3 data set for FAST. The purple line is the power spectral density of pre-fit residuals and the green line is the post-fit residuals after fitting a frequency-dependent and common-mode signal using a 300 days sampling.}
\label{Fig4}
\end{figure}

In order to mitigate the effect of DM variations caused by the interstellar medium, \cite{2013MNRAS.429.2161K} used the simultaneous fitting of a frequency-dependent and frequency-independent function to the residuals.  The function is sampled at specified time intervals. The user is able to choose the time sampling, but they reported optimal time intervals to correct DM variations caused by the interstellar medium.  These optimal time intervals are long and, for the PPTA pulsars, are listed in their Table 2. We have trialled 30, 100, 200 and 300 days sampling. However, as seen from the inset to Figure~1b, the solar wind induces timing residuals on much shorter time scales.  As we expected (and shown in Figure 4), the correction is poor\footnote{Here we only show the results corresponding to the 200 days sampling, but the other sampling intervals give similar results.}. It is clear that the solar wind features in the power spectrum have not been removed and the power spectrum remains orders-of-magnitude higher than the power spectrum of the simulated gravitational wave signal.

The \cite{2013MNRAS.429.2161K} method does not require a common gridding between the frequency-dependent and the frequency-independent functions.  We expect that the frequency-dependent function (representing the solar wind) will vary quickly, but the frequency-independent function (such as the gravitational wave signal) can be sampled on a much sparser grid.  We therefore fitted using a grid spacing of 10 days for the DM (frequency-dependent) function combined with a 300 days sampling for the common mode (frequency-independent) function.

\begin{figure}[h]
\setlength{\belowcaptionskip}{-2.cm}
\centering
\includegraphics[angle=270,width=11cm]{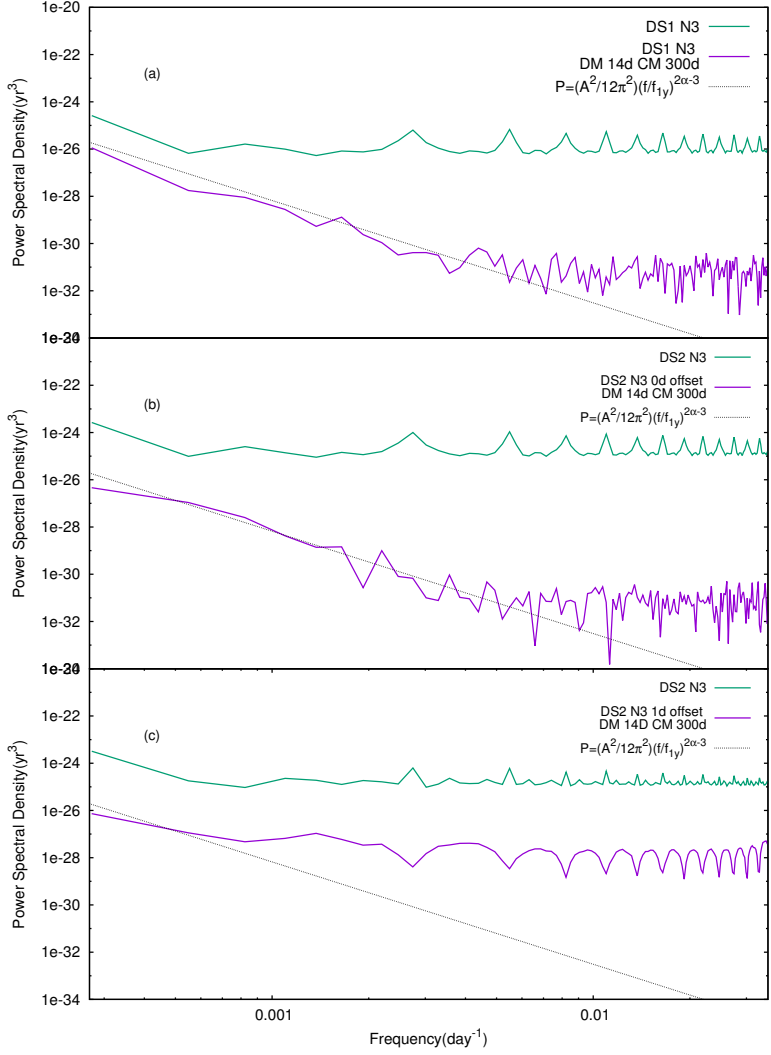}
\caption{Pre-fit (purple) and post-fit (green) power spectral densities for (a) a simulation of FAST, (b) Parkes/MWA without any time offset and (c) Parkes/MWA with a 1 day time offset.}
\label{Fig5}
\end{figure}

We trialled this method on both DS1 and DS2 and present the resulting power spectra in Figure 5. We demonstrate that using a sampling (14 days) for the frequency-dependent function that implies that we can model any DM variations occurring during a single observing epoch.  We simultaneously fit a 300 days frequency-independent function. In Figure 5a we present the pre- and post-fit spectra for our simulation of FAST (DS1)\footnote{Note that the fitting did not converge with a single iteration and as the gravitational wave signal induced residuals significantly larger than the 100\,ns error bars and it was necessary to use the Cholesky fitting routines as described by \cite{2011MNRAS.418..561C} when carrying out the fit.}.  In Figure 5b we show the same, but for the Parkes/MWA data set in which both telescopes are assumed to observe at the same time.  Both panels (a) and (b) indicate that this process works well; we can remove the signature of the solar wind without removing the gravitational wave signal.  However, in Figure 5c we show the resulting power spectra if the Parkes and MWA observations are offset in time by 1 day.  In this case it is clear that we are unable to completely remove the solar wind variations (because the solar wind varies on a time scale faster than a day; see Figure 5 in \citep{2007MNRAS.378..493Y} for a demonstration of this with actual observations) although we note that the lowest frequency channels in the power spectrum are now at the expected level of the gravitational wave signal, also highlight that our simulated gravitational wave signal is at least one-order-of-magnitude higher than the currently expected level.

\section{Discussion and conclusion}

\cite{2016MNRAS.455.4339T} highlighted that when the line-of-sight to pulsars at low ecliptic latitudes passes close to the Sun then the effects of the solar wind can become large. Typically the PTA experiments simply exclude observations within a certain angle from the Sun, but with sufficiently high precision (as will be achievable with telescopes like FAST and the SKA) simply excluding observations will not always be practical.  \cite{2016MNRAS.455.4339T} attempted to model the solar wind using a 100 days gridding, but noted that this was ineffective as the variations caused by the solar wind are narrow, cuspy and change from year to year.  Other correction/modelling methods, such as the Bayesian analysis methods \citep{2014MNRAS.437.3004L} also assume simplistic models for the solar wind.

\begin{table}[h]
\begin{center}
\begin{tabular}{|c|c|c|c|}
\hline
Pulsar & Ecliptic latitude(deg) & Peak amplitude for 270\,MHz & Peak amplitude for 1470\,MHz \\ \hline
J1713$+$0747 & 30.70 & 6.75($\mu$s) & 0.29($\mu$s) \tabularnewline  \hline
J1744$-$1134 & 11.81 & 22.5($\mu$s) &  0.9($\mu$s) \tabularnewline  \hline
J1911$-$1114 & 11.09 & 31.8($\mu$s) & 1.2($\mu$s) \tabularnewline  \hline
J1939$+$2134 & 42.30 & 7.7($\mu$s) & 0.32($\mu$s) \tabularnewline  \hline
J2229$+$2643 & 33.29 & 10.4($\mu$s) & 0.51($\mu$s) \tabularnewline  \hline
\end{tabular}
\caption{A list of pulsars that FAST is expected to time with 100 ns timing precision. The second column is their ecliptic latitude. The third and the fourth columns is the peak amplitude of the solar wind signal in our simulation for the low frequency (270 \,MHz) data and high frequency (1470 \,MHz), respectively.}
\end{center}
\end{table}

Our results indicate that the \cite{2013MNRAS.429.2161K} procedure can be used to model and remove the solar wind effect assuming that the frequency-dependent function has the same gridding as the observing cadence and a wide-bandwidth receiver is used.  This is promising for FAST assuming that the wide-bandwidth receiver is available for high-precision pulsar timing.   As the primary goal for early science operations for FAST is to carry out a survey with a multibeam, 20\,cm receiver, it is possible that the wide-bandwidth receiver will not be commonly available.  In this case it will become much harder to correct for the effect of the solar wind and as shown in the Parkes-MWA simulations it will not be trivial to model the solar wind sufficiently using other telescope observations (unless they are obtained simultaneously with the FAST data).

The sampling required for the solar wind correction is much faster than the optimal sampling given by \cite{2013MNRAS.429.2161K} for modelling and removing DM variations caused by the interstellar medium.  When processing actual PTA data sets it will therefore be necessary to measure the DM variations on an uneven grid spacing.  At times when the line of sight to the pulsar goes close to the Sun then a fast sampling will be required. At other times the \cite{2013MNRAS.429.2161K} sampling will need to be used.  As the solar wind is variable and also the interstellar medium variations can have abrupt changes (for instance, \cite{2015ApJ...808..113C}; reports observations of two extreme scattering events) it is likely that the grid spacing will need to be chosen carefully and manually.

We note that the pulsar we chose in this paper (PSR~J1022$+$1001) is the PTA pulsar that has the smallest ecliptic latitude and therefore the solar wind will have the largest effect for this pulsar. Zhang et al. (in preparation) suggests that five IPTA pulsars (PSRs~J1713$+$0747, J1744$-$1134, J1911$-$1114, J1939$+$2134 and J2229$+$2643) are likely to have ToA uncertainties smaller than 100\,ns during a typical FAST timing observation.  We have re-simulated our daily-sampled FAST data set for these pulsars. Clearly the signal strength amplitude is related to the ecliptic latitude. However, as shown in Table 1, for all pulsars the solar wind effect is detectable in the simulated data sets and for some pulsars (such as PSRs~J1911$-$1114 and J1744$-$1134) the signal is significantly higher than the 100 ns level.

We also note that the amplitude of gravitational waves we set is much larger than our current upper bounds \citep{2015Sci...349.1522S}. To make sure our method is effective for detecting gravitational wave, we re-simulate data for Figure~5b with a lower value of the background amplitude, $A = 10^{-15}$.  The results which is shown in Figure 6 and we note that the algorithm presented here works  well even with the lower gravitational wave background amplitude.

\begin{figure}
\begin{center}
\includegraphics[angle=270,width=11cm]{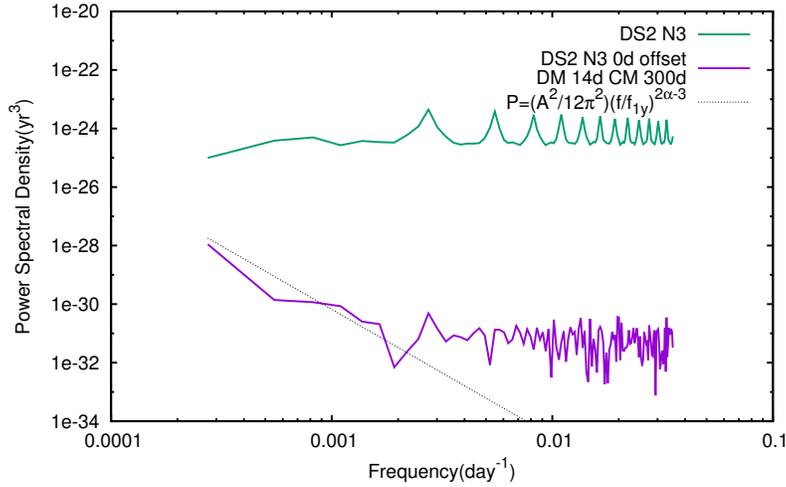}
\end{center}
\caption{Same figure as Figure~5b but change the amplitude of gravitational wave background to $A = 10^{-15}$.}
\end{figure}

For this paper we have studied the use of the \cite{2013MNRAS.429.2161K} method for modelling and removing the effect of the solar wind from pulsar timing array data sets.  Other methods are available.  For instance, the ``DMX" procedure (e.g., \citealt{2013ApJ...762...94D}) can be used in a similar manner, but the common-mode signal must be accounted for in the fitting process to ensure that the gravitational wave signal is not removed.  Similarly, Bayesian methods exist (e.g., \citealt{2014MNRAS.437.3004L}), but producing a model for the solar wind is non-trivial and, to our knowledge, has not been attempted within the Bayesian frameworks.

In conclusion, the solar wind will induce timing residuals that will need to be modelled and/or removed when analysing PTA data.  Our work suggests that this can be done with a wide-enough, simultaneous observing frequency coverage. We also note that the solar wind is intrinsically interesting. Instead of simply removing and discarding the solar wind signature, it can be used (as shown by \citealt{2012MNRAS.422.1160Y}) to improve the accuracy of magnetic field estimates of the solar wind near the sun.

\section{Acknowledgements}

This work is supported by West Light Foundation of CAS(No.XBBS201322) and National Natural Science Foundation of China(Nos.11403086 and U1431107), the Strategic Priority Research Programme(B) of the Chinese Academy of Sciences(No.XDB23010200). The FAST FELLOWSHIP is supported by Special Funding for Advanced Users, budgeted and administrated by Center for Astronomical Mega-Science,  Chinese Academy of Sciences (CAMS). We thank Prof. Zong-hong Zhu for facilitating this project and allowing Ze-xi Niu to spend time in Australia working with the Parkes Pulsar Timing Array project team.  We specifically thank Dr R. Shannon for useful comments on this project. We also thank Dr. Xing-Jiang Zhu for his great help.

\bibliographystyle{raa}
\bibliography{ref.zexi}

\begin{thebibliography}{23}
\providecommand\natexlab[1]{#1}
\providecommand\JournalTitle[1]{#1}

\bibitem[{Arzoumanian} {et~al.}(2016)]{2016ApJ...821...13A}
{Arzoumanian}, Z., {Brazier}, A., {Burke-Spolaor}, S., {et~al.} 2016, \apj,
  821, 13

\bibitem[{Bhat} {et~al.}(2014)]{2014ApJ...791L..32B}
{Bhat}, N.~D.~R., {Ord}, S.~M., {Tremblay}, S.~E., {et~al.} 2014, \apjl, 791,
  L32

\bibitem[{Coles} {et~al.}(2015)]{2015ApJ...808..113C}
{Coles}, W.~A., {Kerr}, M., {Shannon}, R.~M., {et~al.} 2015, \apj, 808, 113

\bibitem[{Coles} {et~al.}(2011)]{2011MNRAS.418..561C}
{Coles}, W., {Hobbs}, G., {Champion}, D.~J., {Manchester}, R.~N., \&
  {Verbiest}, J.~P.~W. 2011, \mnras, 418, 561

\bibitem[{Demorest} {et~al.}(2013)]{2013ApJ...762...94D}
{Demorest}, P.~B., {Ferdman}, R.~D., {Gonzalez}, M.~E., {et~al.} 2013, \apj,
  762, 94

\bibitem[{Desvignes} {et~al.}(2016)]{2016MNRAS.458.3341D}
{Desvignes}, G., {Caballero}, R.~N., {Lentati}, L., {et~al.} 2016, \mnras, 458,
  3341

\bibitem[{Dewdney} {et~al.}(2009)]{2009IEEEP..97.1482D}
{Dewdney}, P.~E., {Hall}, P.~J., {Schilizzi}, R.~T., \& {Lazio}, T.~J.~L.~W.
  2009, IEEE Proceedings, 97, 1482

\bibitem[{Hobbs} {et~al.}(2006)]{2006MNRAS.369..655H}
{Hobbs}, G.~B., {Edwards}, R.~T., \& {Manchester}, R.~N. 2006, \mnras, 369, 655

\bibitem[{Hobbs} {et~al.}(2004)]{2004MNRAS.353.1311H}
{Hobbs}, G., {Lyne}, A.~G., {Kramer}, M., {Martin}, C.~E., \& {Jordan}, C.
  2004, \mnras, 353, 1311

\bibitem[{Hobbs} {et~al.}(2009)]{2009MNRAS.394.1945H}
{Hobbs}, G., {Jenet}, F., {Lee}, K.~J., {et~al.} 2009, \mnras, 394, 1945

\bibitem[{Keith} {et~al.}(2013)]{2013MNRAS.429.2161K}
{Keith}, M.~J., {Coles}, W., {Shannon}, R.~M., {et~al.} 2013, \mnras, 429, 2161

\bibitem[{Lam} {et~al.}(2016)]{2016ApJ...821...66L}
{Lam}, M.~T., {Cordes}, J.~M., {Chatterjee}, S., {et~al.} 2016, \apj, 821, 66

\bibitem[{Lentati} {et~al.}(2014)]{2014MNRAS.437.3004L}
{Lentati}, L., {Alexander}, P., {Hobson}, M.~P., {et~al.} 2014, \mnras, 437,
  3004

\bibitem[{Loi} {et~al.}(2015)]{2015RaSc...50..574L}
{Loi}, S.~T., {Trott}, C.~M., {Murphy}, T., {et~al.} 2015, Radio Science, 50,
  574

\bibitem[{Manchester} {et~al.}(2013)]{2013PASA...30...17M}
{Manchester}, R.~N., {Hobbs}, G., {Bailes}, M., {et~al.} 2013, \pasa, 30, e017

\bibitem[{Reardon} {et~al.}(2016)]{2016MNRAS.455.1751R}
{Reardon}, D.~J., {Hobbs}, G., {Coles}, W., {et~al.} 2016, \mnras, 455, 1751

\bibitem[{Shannon} {et~al.}(2015)]{2015Sci...349.1522S}
{Shannon}, R.~M., {Ravi}, V., {Lentati}, L.~T., {et~al.} 2015, Science, 349,
  1522

\bibitem[{Tiburzi} {et~al.}(2016)]{2016MNRAS.455.4339T}
{Tiburzi}, C., {Hobbs}, G., {Kerr}, M., {et~al.} 2016, \mnras, 455, 4339

\bibitem[{Tingay} {et~al.}(2013)]{2013PASA...30....7T}
{Tingay}, S.~J., {Goeke}, R., {Bowman}, J.~D., {et~al.} 2013, \pasa, 30, e007

\bibitem[{You} {et~al.}(2012)]{2012MNRAS.422.1160Y}
{You}, X.~P., {Coles}, W.~A., {Hobbs}, G.~B., \& {Manchester}, R.~N. 2012,
  \mnras, 422, 1160

\bibitem[{You} {et~al.}(2007{\natexlab{a}})]{2007ApJ...671..907Y}
{You}, X.~P., {Hobbs}, G.~B., {Coles}, W.~A., {Manchester}, R.~N., \& {Han},
  J.~L. 2007{\natexlab{a}}, \apj, 671, 907

\bibitem[{You} {et~al.}(2007{\natexlab{b}})]{2007MNRAS.378..493Y}
{You}, X.~P., {Hobbs}, G., {Coles}, W.~A., {et~al.} 2007{\natexlab{b}}, \mnras,
  378, 493

\bibitem[{Zhang} {et~al.}(2016)]{2016RAA....16..151Z}
{Zhang}, L., {Hobbs}, G., {Li}, D., {et~al.} 2016, Research in Astronomy and
  Astrophysics, 16, 151

\end{thebibliography}

\end{document}